# Holistic Fault Detection and Diagnosis System in Imbalanced, Scarce, Multi-Domain (ISMD) Data Setting for Component-Level Prognostics and Health Management (PHM)

Ali Rohan[*]

*Department of Mechanical, Robotics, and Energy Engineering, Dongguk University, 30 Pildong 1 Gil, Junggu, 04620, Seoul, South Korea*

*Faculty of Medicine and Health Sciences, University of Nottingham, Sutton Bonington, LE12 5RD, Loughborough, United Kingdom*

## Abstract

In the current Industrial 4.0 revolution, Prognostics and Health Management (PHM) is an emerging field of research. The difficulty of obtaining data from electromechanical systems in an industrial setting increases proportionally with the scale and accessibility of the automated industry, resulting in a less interpolated PHM system. To put it another way, the development of an accurate PHM system for each industrial system necessitates a unique dataset acquired under specified conditions. In most circumstances, obtaining this one-of-a-kind dataset is difficult, and the resulting dataset has a significant imbalance, a lack of certain useful information, and multi-domain knowledge. To address this, this paper provides a fault detection and diagnosis system that evaluates and pre-processes Imbalanced, Scarce, Multi-Domain (ISMD) data acquired from an industrial robot utilizing Signal Processing (SP) techniques and Deep Learning-based (DL) domain knowledge transfer. The domain knowledge transfer is used to produce a synthetic dataset with a high interpolation rate that contains all the useful information about each domain. For domain knowledge transfer and data generation, Continuous Wavelet Transform (CWT) with Generative Adversarial Network (GAN) was used, as well as Convolutional Neural Network (CNN) to test the suggested methodology using transfer learning and categorize several faults. The proposed methodology was tested on a real experimental bench that included an industrial robot created by Hyundai Robotics Co. This development resulted in a satisfactory resolution with 99.7% (highest) classification accuracy achieved by transfer learning on several CNN benchmark models.

---

[*]Corresponding author.
*Email address:* `ali_rohan2003@hotmail.com; alirohan@dongguk.edu;`

## 1. Introduction

In contemporary industrial settings, the majority of tasks are performed by electromechanical devices, such as robots. These robots are made up of several electrical and mechanical components joined together to perform a uniquely engineered operation. These electromechanical components are vulnerable to degradation, due to continued operation. Over time, proper assessment and maintenance strategies are required to prevent fatal damage. To counter this, Prognostics and Health Management (PHM) has evolved as an attractive method in establishing techniques for system health monitoring, diagnostics, Remaining Useful Life (RUL) prediction, and prognostics. PHM is considered to be an effective approach to providing comprehensive, tailored solutions for health management [1]. PHM has three critical tasks: 1) fault detection: detection of fault trigger at the early stage of the component or system degradation; 2) fault diagnosis: segregation and identification of fault and its source; and 3) predictions: RUL forecasting. Figure 1 shows the essential tasks involved in a PHM system. PHM can be applied at the component level, system level, or both. PHM at the component level directs the development of health monitoring strategies for specific components, such as electric motors, electronic devices, bearings, and gear reducers. It determines whether the health of the monitored component is time-degraded due to various environmental, operational, and performance-related parameters [2, 3]. In contrast, PHM at the system level evaluates detailed system health, factoring system operation, design, and process-related parameters [4].

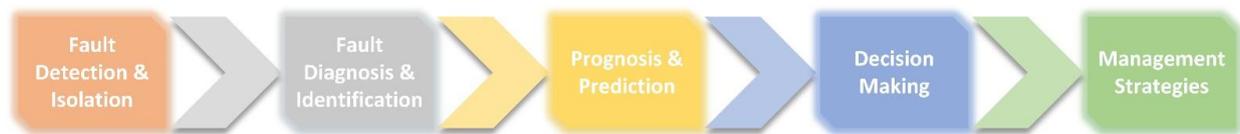

Figure 1: Basic tasks involved in a typical PHM system.

In recent years, the tremendous progress in Artificial Intelligence (AI) has strengthened the potential for designing PHM systems that are powerful enough to detect, diagnose, and predict faults at an earlier stage with high precision. Deep Learning (DL) and Machine Learning (ML) have become essential tools in establishing the decision-making capabilities of a PHM system. Numerous studies have been conducted on designing and implementing such a system at the component and system levels. PHM is generally classified into either mathematical-model or data-driven-based approaches [5–7]. Awareness of the core knowledge of the component understudies, such as material characteristics and architectural attributes, is part of the mathematical model approaches [8, 9], while data-driven approaches derive information from statistical data to forecast the component's health [10–14]. The mathematical model-based methods [15–17] require the development of a physics-based model of an element or system beforehand to develop some PHM strategies and struggle

ali.rohan@nottingham.ac.uk (Ali Rohan)

with several issues, such as a noisy and dynamic work environment restricting the requisite precision for the creation of the desired model.

Furthermore, these models cannot be upgraded by the newly recorded data in real-time. In contrast to the model-based approaches, data-driven approaches [18, 19] are becoming popular, due to their ability to update and transform in real-time under different scenarios. Also, there have been substantial improvements in the computing capabilities of devices with improved sensing technologies that allow efficient data acquisition. Current data-driven approaches require significant information and additional real-time measurements, such as vibration, acoustic emission, laser displacement, temperature, speed, and electrical current [20, 21], to design a PHM system. Recently, researchers have suggested data-driven approaches that focus primarily on DL-based fault diagnosis or prognosis [22, 23]. In contrast, others concentrate on the applicability of a specific item, such as a bearing, or an electronic system [24–26]. Other studies have addressed alternative propositions, such as intelligent condition-based monitoring of Rotating Electrical Machines (REMs) using a sparse auto-encoder approach [27], Rolling Element Bearing (REB) PHM based on Deep Convolutional Neural Network (DCNN) [28], and improved DCNN, i.e., the hierarchical adaptive DCNN [29]. CNN-based mechanical bearing fault detection [30] was introduced as a feature-learning basis for health monitoring to freely learn useful features from the data. Meanwhile, the previous study [31] introduced an ML-based fault detection and diagnostic method based on a different feature selection, extraction, and infusion process.

The success of the aforementioned approaches is highly dependent on the following factors: 1) data availability: if data is available, or can be acquired for specific components or systems; 2) data type: what type of data is known, or can be obtained, such as vibration, acoustic emission, or electric current; 3) data quality: if data is recorded with precision, and is constitutive of all the information required to analyze the features and behaviour of a particular component or system; and 4) data quantity: if data are sufficient in quantity for analysis and the creation of an interpolated PHM system. Generally, due to the complexities of industrial parts and processes, it is difficult to collect data that satisfy all these factors. The raw data collected is often Imbalanced: the numbers of samples per class are not evenly distributed (sometimes a significant volume of data for one class, alluded to as the majority class, and much fewer samples for one or two other classes, alluded to as the minority classes); Scarce: data is not adequate to create sustainable DL models that can be applied at the production stage; and Multi-domain: the dataset is comprised of information regarding several domains, e.g., speed characteristics of a rotating electrical or mechanical component at a different level of speed. Thus, Imbalanced, Scarce, Multi-Domain (ISMD) data mark a significant bottleneck in the growth of PHM systems.

Research continues to tackle the challenges related to DL application for PHM. Previously, in [32], the authors provide a brief introduction to several deep learning models by reviewing and analyzing applications of fault detection, diagnosis, and prognosis. To resolve ISMD data, which poses a significant challenge for AI-based applications. Several proposed methods, such as Data Augmentation and Transfer Learning, have been used to address the imbalance and scarce data problems. The Generative Adversarial Network (GAN) was frequently used for this reason. In [33], authors proposed Balancing-GAN (BAGAN) as an augmentation method for restoring balance in imbalanced datasets, while in [34], a concept for learning the discriminative classifier from unlabeled or partly labelled data using Categorical-GAN (CatGAN) was established. In another variation, a data augmentation process was proposed [35], producing artificial medical images, using GAN to classify liver lesions. Apart from these methodologies, to address current issues with traditional electromechanical system monitoring approaches used in industrial settings, some researchers focused on the Incremental Learning and Novelty Detection methodologies [36, 37]. These methodologies mainly focused on the development of fault detection systems in industrial settings with a lack of data for faulty conditions. Introducing a systematic approach from feature selection, extraction, classification to retraining to include new patterns to the novelty detection and fault identification models, these techniques showed promising results. In recent studies [38, 39], GAN was used for multi-domain image translation with various image syntheses. The architecture is called StarGAN. The study implemented a rigorous approach to convert images from two different domains into a single domain, without losing any useful features, thus leading a pathway for the introduction of domain knowledge transfer using GANs in the field of DL.

ISMD data have been a specific research challenge to creating interpolated DL models, regardless of the type of classification issue. In particular, the problem of multi-domain data has not yet been adequately investigated. Therefore, this specific work expands the ISMD data issue by focusing on the multi-domain data, intending to develop holistic fault detection and diagnosis system for component-level PHM. By adding multiple faults related to a mechanical component, the Rotate Vector (RV) reducer, data in real-time from an industrial robot was recorded for several domains. The recorded data was then pre-processed using signal processing techniques, such as Discrete Wavelet Transform (DWT), and Continuous Wavelet Transform (CWT). The power of StarGAN was used to transfer domain knowledge using image-to-image translation to generate a synthetic dataset with the real dataset. The final dataset was then validated using transfer learning with a variety of CNN benchmark models. As relevant literature has been examined in this study, such a technique has not yet been applied for any practical application, specifically for PHM. Subsequently, the prospects of using Motor Current Signature Analysis (MCSA) to detect and diagnose mechanical faults are also investigated in this study. Earlier, mechanical faults were detected by contrasting approaches, which were vibration signal analysis, acoustic emission, or ferrography analysis. Among these, the most effective way has been vibration analysis. This method faces certain challenges such as its usage requires costly vibration sensors, challenging to place and install in particular areas to record vibration signals. The ambient environment also generates noise, causing erroneous sensor readings. As an option, MCSA has a range of benefits over vibration analysis. MCSA uses the built-in current signal of the motor control unit, having no extra sensors, resulting in a low-cost and less complicated framework. In addition, the current signals are peculiar, being not easily affected by the ambient operating conditions.

The following sections present the details of the proposed fault detection and diagnosis approach: Section 2 defines the materials and methods, including the experimental test bench and the descriptions of the suggested technique; Section 3 presents the results and discussions; and Section 4 concludes the paper.

## 2. Materials and Methods

*2.1. Experimental Test Bench.*

The experimental test bench in this study as shown in Fig. 2. consists of three main components: industrial robot, controller, and Personal Computer (PC). Manufactured by Hyundai Robotics Co., the robot used is modelled as YS080 with a maximum payload capacity of 80 kgf. Figure 3 shows details of the robot as follows: (a) the free body diagram and (b) the Hyundai Robot YS080. The robot comprises of six axes or joints where the individual axis is mounted with an electric motor of diverse specifications, enabling it to move freely 360 degrees about each axis.

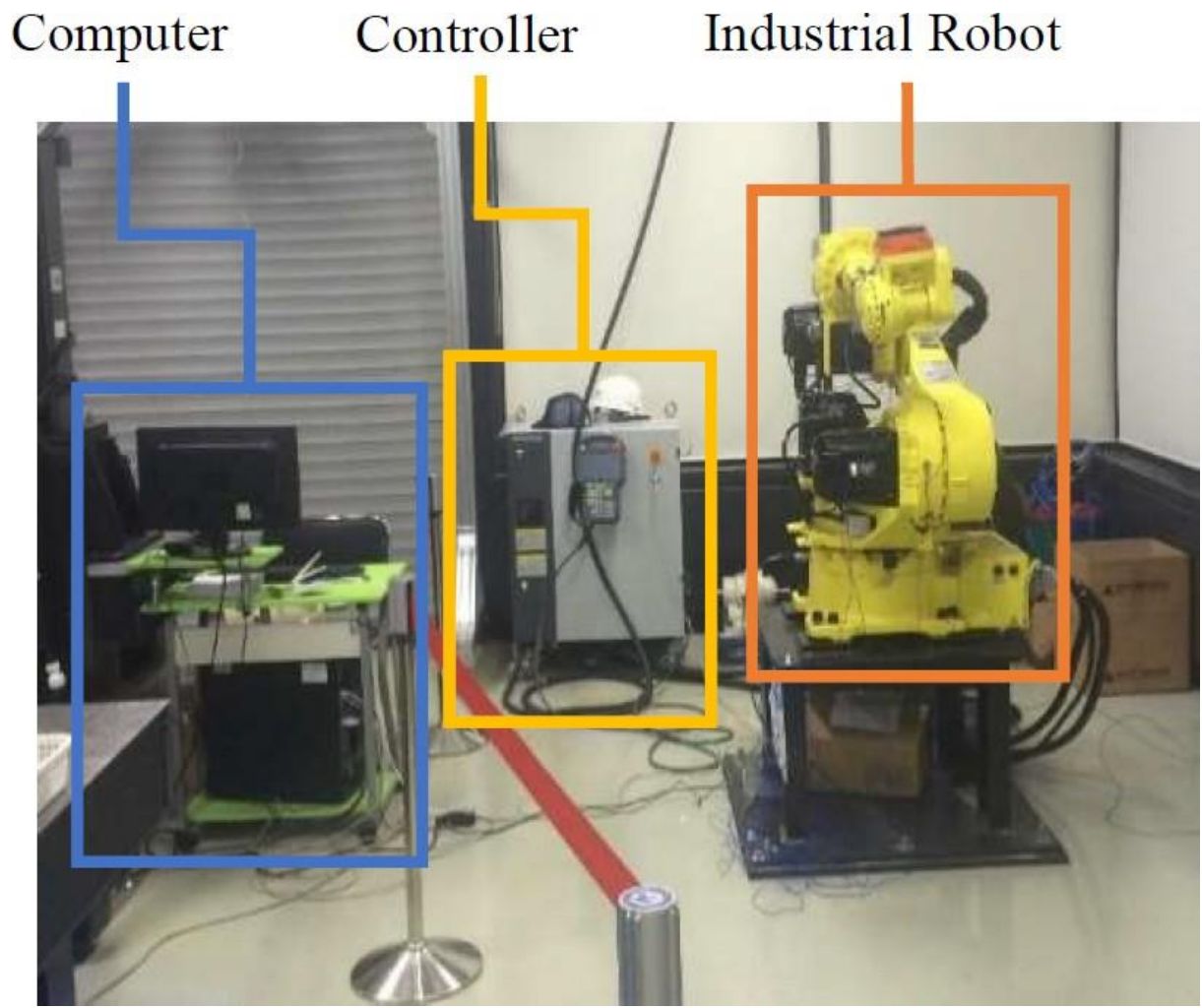

Figure 2: Experimental test bench.

The motors are attached to the reducers at each axis to increase or decrease rotation speed. The robot is operated by sending commands to the controller through a PC, which, in turn, operates electric motors to produce a specific motion. Three-phase servo motors are used on each axis. The power of the motors is configured based on the amount of mechanical load on each axis. The first three axes are equipped with high specification motors, whereas the others are equipped with lower specifications. Table 1 summarizes the details of electric motors.

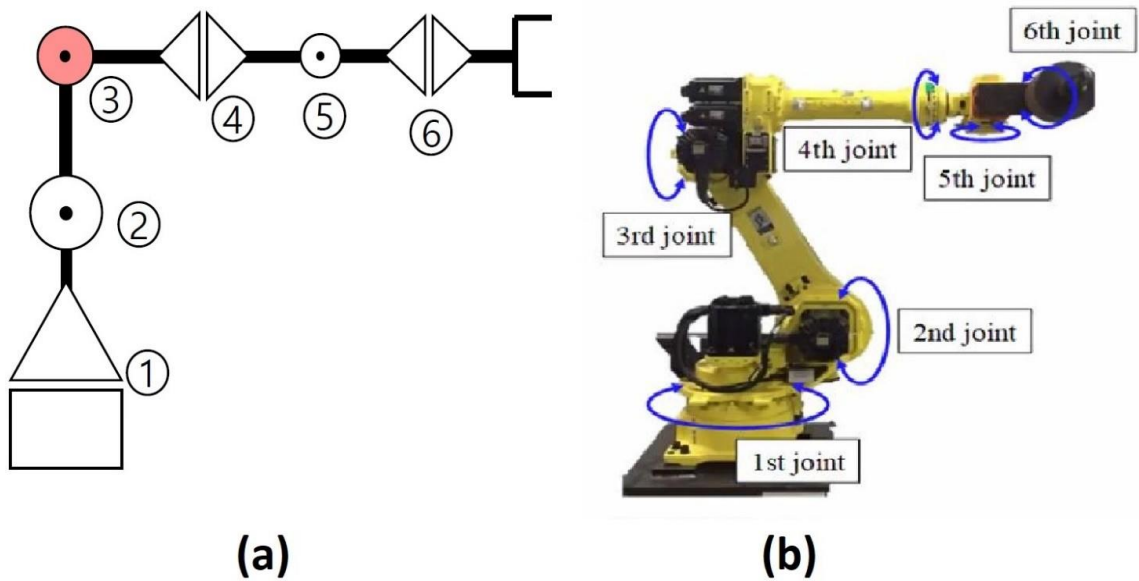

(a)         (b)

Figure 3: a) Free body diagram, and (b) Hyundai Robot YS080.

Table 1: Specifications of the electric motors.

| Axes No. | Power (kW) | Speed (rpm) | Voltage (V) | Current (A) | Frequency (Hz) |
|---|---|---|---|---|---|
| 1, 2, 3 | 5.9 | 2,000 | 200 | 25.1 | 166 |
| 4, 5, 6 | 2 | 3,000 | 200 | 11.7 | 250 |

## 2.2. The Architecture of the Proposed Methodology

Figure 4 shows the overall basic architecture of the proposed methodology for fault detection and diagnosis system. The proposed method is divided into three steps:

1. Data Analysis
2. Domain Knowledge Transfer
3. Data Split and Classification

Several experiments were conducted on the experimental test bench shown in Figs. 2 & 3 under different conditions to record the data. The recorded ISMD dataset was further used to create a multi-domain infused dataset using GAN and image-to-image translation. Finally, the generated dataset and original ISMD dataset were used, and a CNN was used to perform the classification task between several faults related to the RV reducer. The following subsection contains a detailed explanation of the proposed methodology.

### 2.2.1. Data Analysis

*Data Acquisition.* The three-phase current signal data for each axis servo motor was recorded using Hall Effect Base Linear Current Sensors WCS6800. Sensors were installed at each phase of the electric motor on each axis of the robot. The current signals were recorded using a total of 18 current sensors for 6 motors. Figure 5 describes the data acquisition system. National Instrument NI DAQ 9230 modules were used to acquire the data. This module for data recording sends the collected signal to a PC with a LabView program installed on it. The received data were processed, and a concluding archive was established containing the signal information for the three-phase current signals of each axis motor.

The data were recorded concurrently for each motor under various fault situations. An RV reducer eccentric bearing fault was introduced into the reducer coupled with the $4^{th}$ axis motor in one case. In another case, the fault was introduced by replacing the RV reducer with a degraded one. The data was recorded for a total of three classes: normal, faulty (RV reducer eccentric bearing fault), and faulty age (RV reducer ageing fault). Figure 6 pinpoints the position of the faults in the Hyundai Robot with a comprehensive logical view. Figure 7 depicts the fault modes using an example of a fault specimen. For several cycles, the robot was operated to move freely around the axis of rotation.

Data-driven methods typically involve a huge volume of data and more samples, which is the optimal case. However, in this case, the data was recorded for 10 cycles for each axis to generate ISMD data. Each cycle corresponds to the completion of motion along a particular axis. To introduce the scarcity into the dataset, 10 cycles were chosen. The data were recorded with an imbalance between the three classes (each class containing a different number of samples). Subsequently, the motors were operated at various speed profiles ranging from 10 to 100% of the rated speed, to obtain the data for multi-domain. Each speed level is characterized as a single domain since each speed includes information about the various time and frequency components of the original signal. Figure 8 shows the details of the hardware used in the process of data acquisition. For each axis motor, the data were recorded, although the fault is only placed into the RV reducer at axis 4, as due to mechanical coupling, a fault in one axis could influence the operation and performance of the other axis motors. Table 2 shows the details of the recorded dataset. There are multiple speed domains from 10 to 100%. The data are imbalanced with a different number of samples for each class, such as 30/27/24 for Normal, Faulty, and Faulty Age, respectively. The data are scarce, as only 30 samples are recorded for the single domain at one axis. The final dataset consisted of 4,860 samples for the 3 classes.

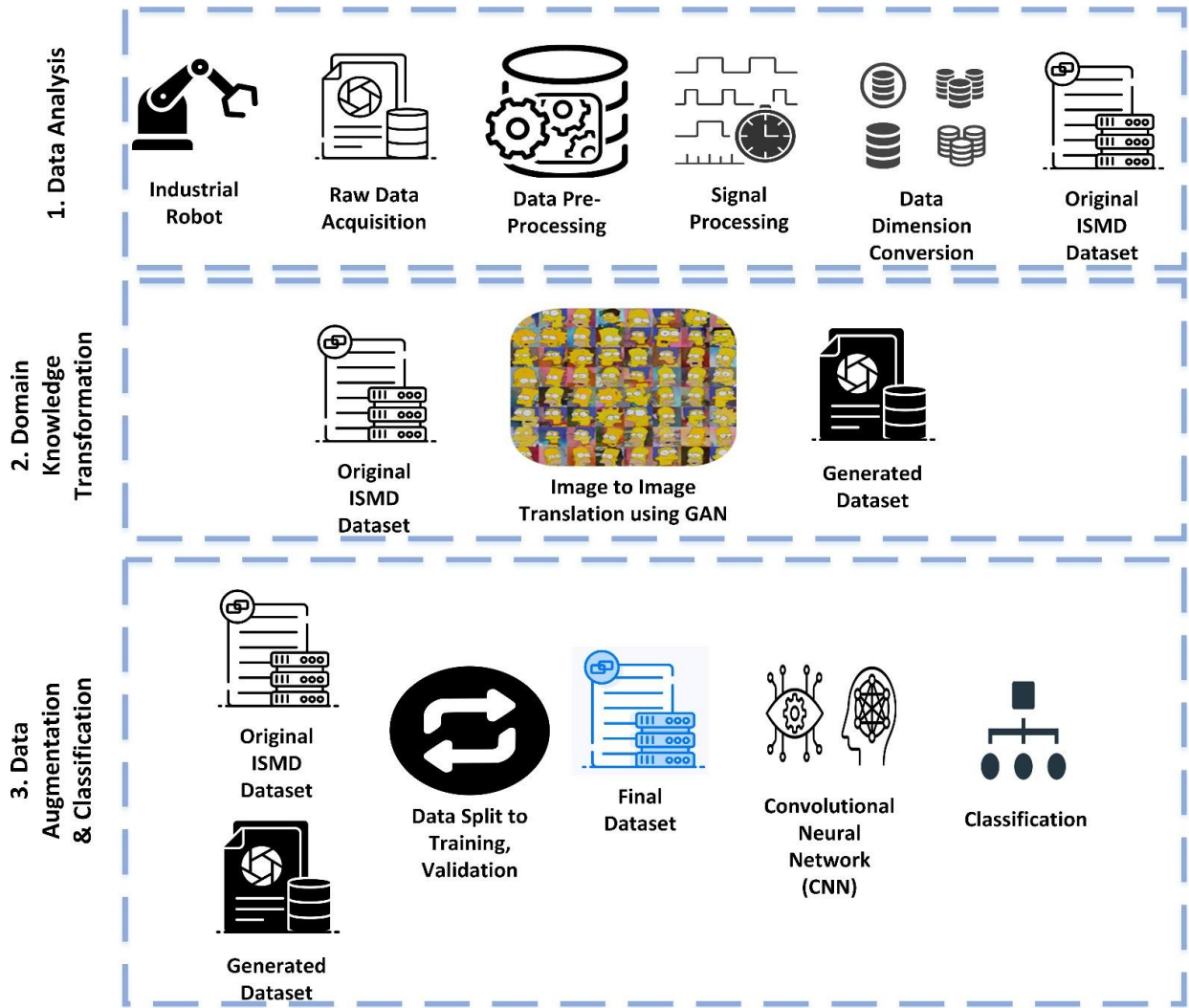

Figure 4: Basic architecture of the proposed fault detection and diagnosis system.

*Data Pre-processing.* To reduce the dimensionality and size of the recorded dataset, DQ0 Transformation was implemented on the three-phase current signals. The DQ0 transformation was utilized to convert the three-phase current signal to a two-phase current signal in such a way the information in the remaining two signals was preserved. The DQ0 transformation is a well-known technique to reduce the dimensions of the electric signal, transforming a three-phase current signal to an arbitrary rotating framework of DQ0 by projecting the knowledge from a three-dimensional to a two-dimensional space. The resulting signals can be described by a circle in the projected two-dimensional space. This helps to simplify the frequency analysis, since the circle correlates to the signal, and helps in conserving the magnitude of the current or voltage signals. In this work, the sinusoidal-based DQ0 Transformation, which is given in Eq. (1) was used. Figure 9 shows the three-phase and two-dimensional representation of the DQ0 transformation.

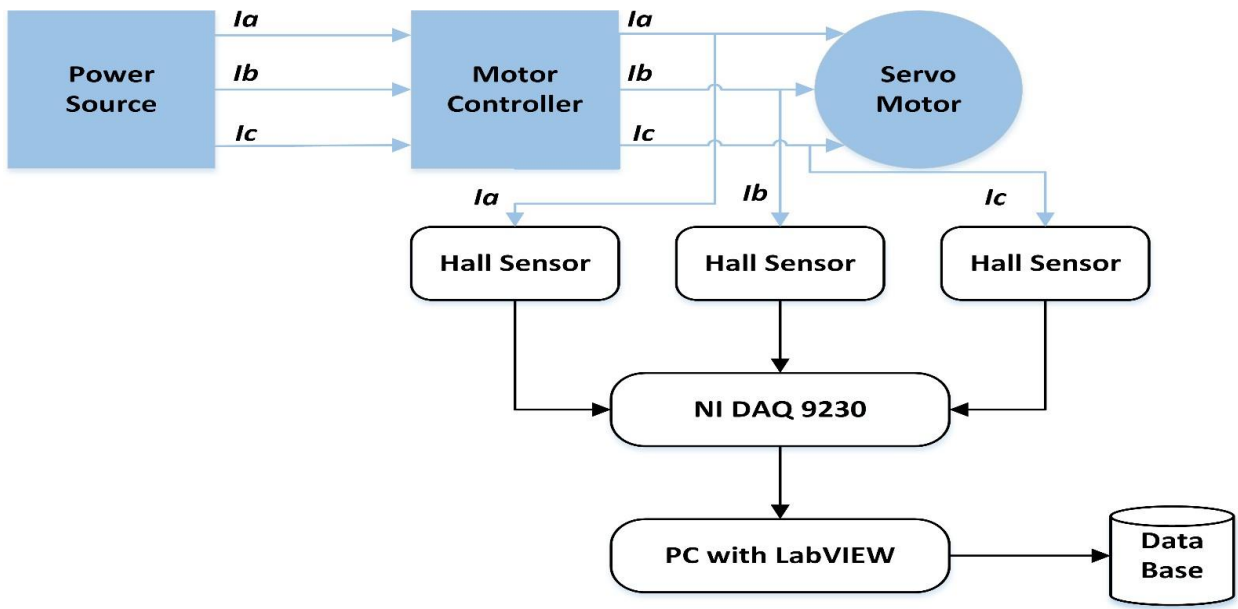

Figure 5: Block diagram of the data collection process.

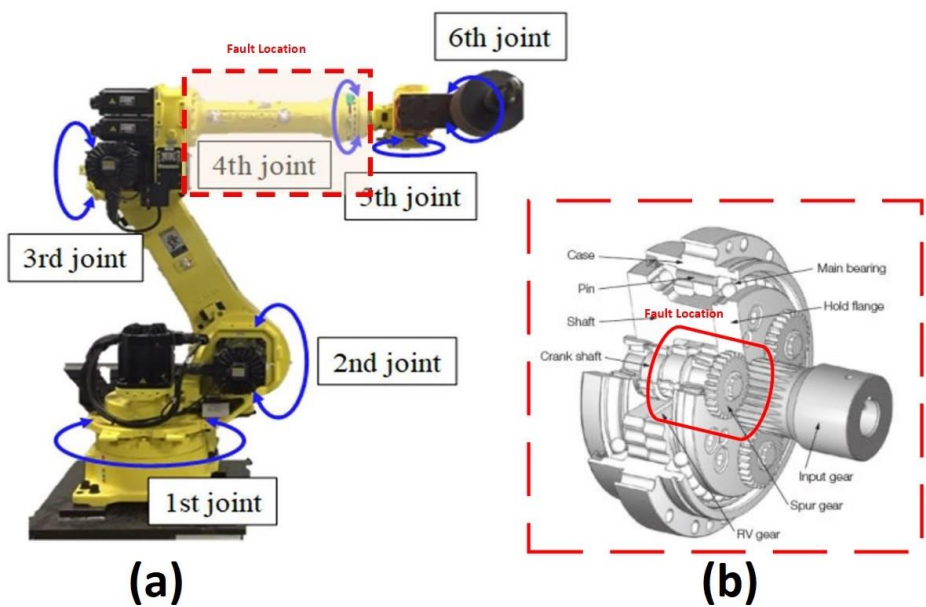

Figure 6: (a) Location of the fault, and (b) comprehensive logical view.

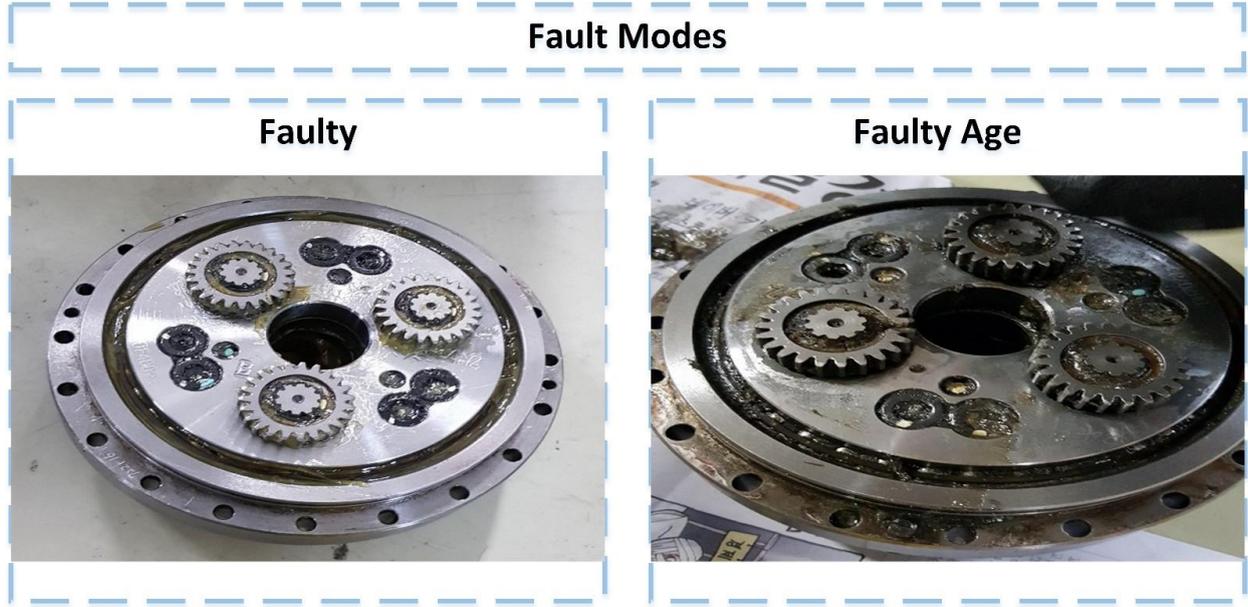

Figure 7: An example of a faulty specimen and fault types.

$$T_{abc-dq} = \sqrt{\frac{2}{3}} \begin{bmatrix} \sin \omega t & \sin \left( \omega t - \dfrac{2\pi}{3} \right) & \sin \left( \omega t + \dfrac{2\pi}{3} \right) \\ \cos \omega t & \cos \left( \omega t - \dfrac{2\pi}{3} \right) & \cos \left( \omega t + \dfrac{2\pi}{3} \right) \\ \dfrac{1}{\sqrt{2}} & \dfrac{1}{\sqrt{2}} & \dfrac{1}{\sqrt{2}} \end{bmatrix} \tag{1}$$

*Signal Processing and Data Dimension Conversion .* Signals extracted from the sensors are in raw form, and normally require pre-processing to eliminate noise and unnecessary information. Signal processing techniques are used for this purpose, which helps in analyzing the features of a specific signal in time, frequency, and time-frequency domains. There are several types of signal processing techniques based on different domains. These techniques are categorized as time-domain, frequency-domain, and time-frequency domain analysis. Statistical parameters illustrating valuable knowledge in the time-domain are extracted from the signal [40] [41], [42] in the time-domain analysis. Whereas for frequency-domain analysis, Fourier transform (FT) has been the most frequently used method; it decomposes the signals into constituent frequencies. Another technique is the Fast-Fourier Transform (FFT), which is widely used for the analysis of continuous-time signals. This transformation uses the spectral frequencies for the analysis of a signal. However, for the processing of non-stationary signals, such as this particular work, time- and frequency-domain analysis techniques have certain limitations.

Time-frequency domain analysis, a combination of frequency and time domains, has been established to improve these limitations [43] . The standard method used for this purpose is known as the Short-Time Fourier Transform (STFT) [44] , which segregates the whole signal via FT and windows of small time-periods into different segments. Another common method with a similar purpose is the Wavelet Transform (WT) [45, 46]. WT is an empirical technique that utilizes the interpretation of wavelet decomposition as a scaling idea for time-varying or non-stationary signals. This is a better approach for analyzing signals with a dynamic frequency spectrum with high resolution in both the time- and frequency-domains. Unlike other techniques, it not only informs which frequencies are present in a signal, but also at which time these frequencies have occurred.

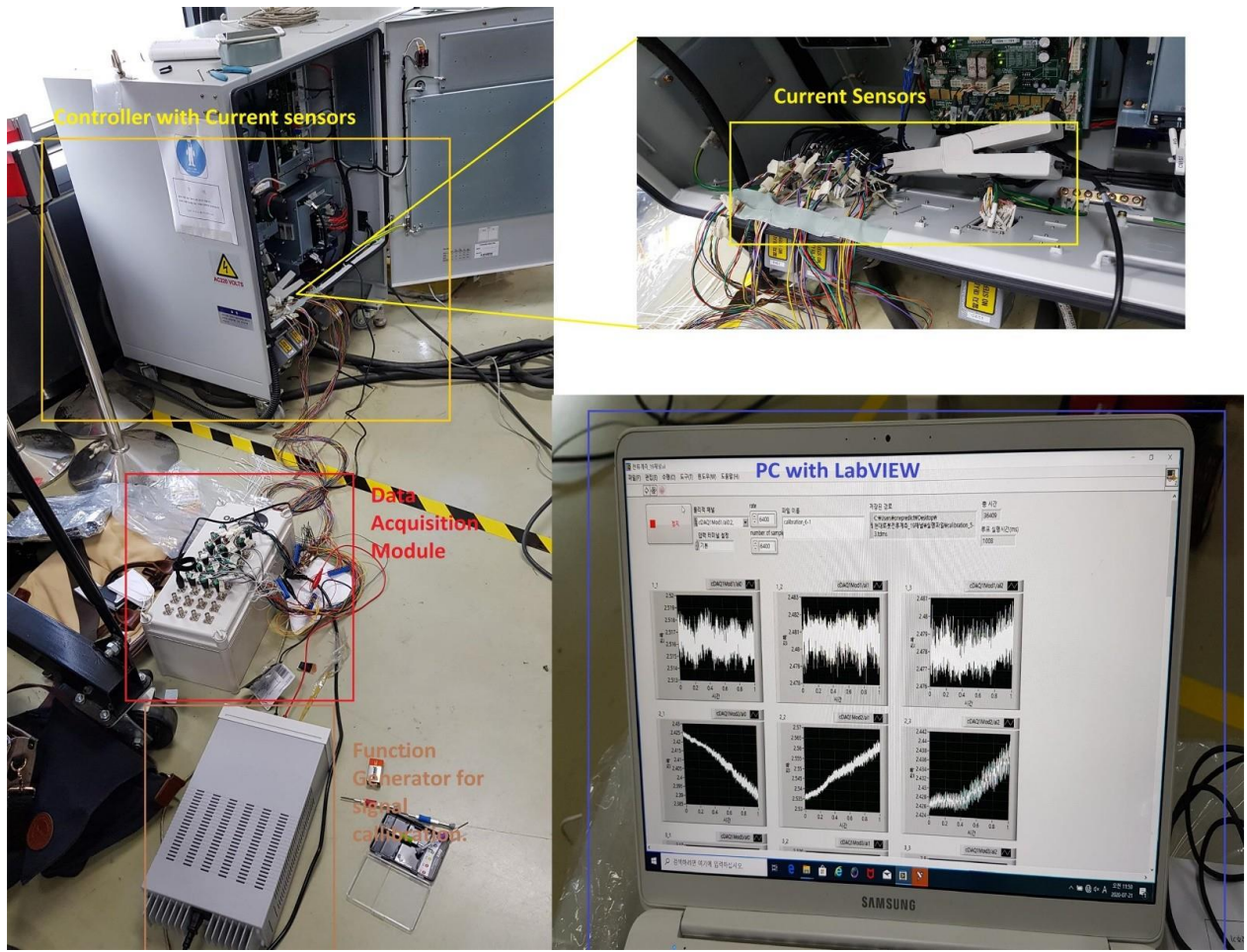

Figure 8: Details of the hardware used in the data collection.

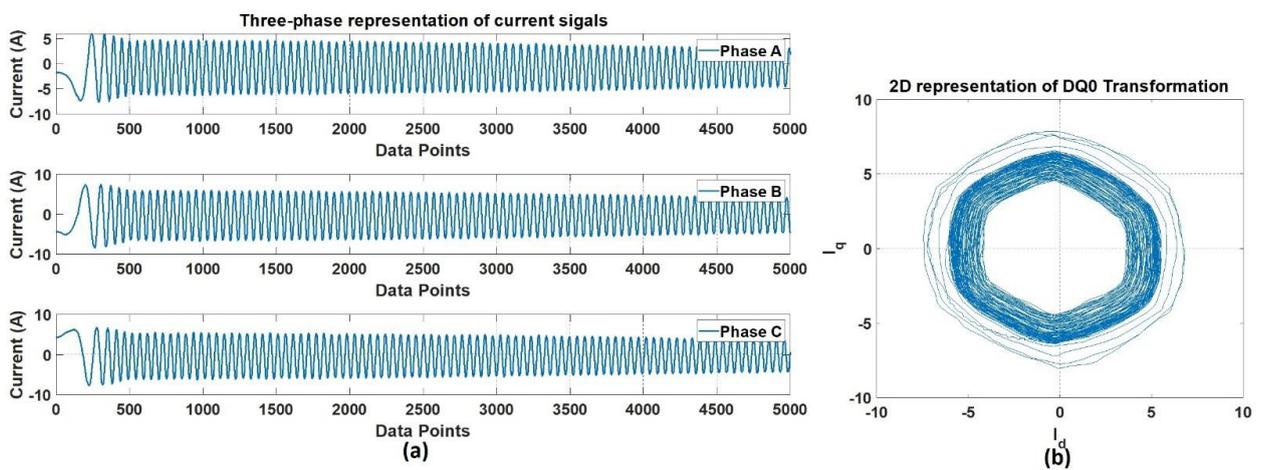

Figure 9: Three-phase current signal with DQ0 representation.

Table 2: Details of the recorded dataset.

| Recorded Dataset | Multi-Domain (Speed Profiles) | | | | | | | | | |
|---|---|---|---|---|---|---|---|---|---|---|
| | 10% | 20% | 30% | 40% | 50% | 60% | 70% | 80% | 90% | 100% |
| | Class; Normal \ Faulty \ Faulty Age, Number of Samples (cycles) | | | | | | | | | |
| Axis 1 | 30 \ 27 \ 24 | 30 \ 27 \ 24 | 30 \ 27 \ 24 | 30 \ 27 \ 24 | 30 \ 27 \ 24 | 30 \ 27 \ 24 | 30 \ 27 \ 24 | 30 \ 27 \ 24 | 30 \ 27 \ 24 | 30 \ 27 \ 24 |
| Axis 2 | 30 \ 27 \ 24 | 30 \ 27 \ 24 | 30 \ 27 \ 24 | 30 \ 27 \ 24 | 30 \ 27 \ 24 | 30 \ 27 \ 24 | 30 \ 27 \ 24 | 30 \ 27 \ 24 | 30 \ 27 \ 24 | 30 \ 27 \ 24 |
| Axis 3 | 30 \ 27 \ 24 | 30 \ 27 \ 24 | 30 \ 27 \ 24 | 30 \ 27 \ 24 | 30 \ 27 \ 24 | 30 \ 27 \ 24 | 30 \ 27 \ 24 | 30 \ 27 \ 24 | 30 \ 27 \ 24 | 30 \ 27 \ 24 |
| Axis 4 | 30 \ 27 \ 24 | 30 \ 27 \ 24 | 30 \ 27 \ 24 | 30 \ 27 \ 24 | 30 \ 27 \ 24 | 30 \ 27 \ 24 | 30 \ 27 \ 24 | 30 \ 27 \ 24 | 30 \ 27 \ 24 | 30 \ 27 \ 24 |
| Axis 5 | 30 \ 27 \ 24 | 30 \ 27 \ 24 | 30 \ 27 \ 24 | 30 \ 27 \ 24 | 30 \ 27 \ 24 | 30 \ 27 \ 24 | 30 \ 27 \ 24 | 30 \ 27 \ 24 | 30 \ 27 \ 24 | 30 \ 27 \ 24 |
| Axis 6 | 30 \ 27 \ 24 | 30 \ 27 \ 24 | 30 \ 27 \ 24 | 30 \ 27 \ 24 | 30 \ 27 \ 24 | 30 \ 27 \ 24 | 30 \ 27 \ 24 | 30 \ 27 \ 24 | 30 \ 27 \ 24 | 30 \ 27 \ 24 |
| Total No. of Samples | 180 \ 162 \ 144 | 180 \ 162 \ 144 | 180 \ 162 \ 144 | 180 \ 162 \ 144 | 180 \ 162 \ 144 | 180 \ 162 \ 144 | 180 \ 162 \ 144 | 180 \ 162 \ 144 | 180 \ 162 \ 144 | 180 \ 162 \ 144 |
| Dataset Size | 4860 X 3 (No. of Samples X No. of Classes) | | | | | | | | | |

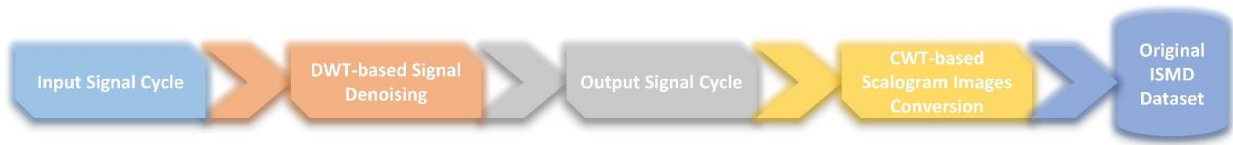

Figure 10: The flow of signal processing and data dimension conversion technique.

The WT is mainly divided into two types: Discrete Wavelet Transform (DWT) and Continuous Wavelet Transform (CWT). Each of these types has its function in signal analysis. DWT is a non-redundant transformation that mainly focuses on one-to-one interaction between the information in the signal domain and the transform domain. This close interaction makes the DWT more appropriate for applications such as signal de-noising and reconstruction. However, the CWT is suitable for scalograms, because the analysis window can be sized and configured at any position. This adaptability facilitates the generation of smooth images. Therefore, in this work, to construct a dataset that was rich with features of both times- and frequency- domains, the DWT was used for signal de-noising, and the CWT for the generation of scalogram representations of each cycle for multiple speed domains. The CWT was used for the conversion of data from a one-dimensional signal to a two-dimensional image. The CWT can be mathematically given as:

$$X_{\omega}(a,b) = \frac{1}{|a|^{\frac{1}{2}}} \int_{-\infty}^{\infty} x(t) \overline{\psi}\left(\frac{t-b}{a}\right) dt \qquad (2)$$

Where, $\psi(t)$ is the continuous mother wavelet, which gets scaled by a factor of *a* and translated by a factor of *b*. Figure 10 shows the overall flow of the signal processing and data dimension conversion techniques implemented in this work. Figure 11 shows an example of the de-noised signal using DWT where the blue signal is the original signal, and the green signal is the de-noised signal. The signal is of one cycle of mechanical rotation of the robot along one axis, rather than an electrical cycle. On the other hand, Fig. 12 shows an example of the scalogram images after the implementation of CWT on the de-noised cycles for a single domain of 10 % rotation speed.

*Original ISMD data* . The final dataset after preprocessing and signal processing were stored in a database of scalogram images reflecting the original current signal cycles in various speed domains. As stated earlier, the dataset in this work focuses on a single parameter of speed to construct multi-domain data. Preliminary findings and signal processing aided the simplification of the data collection by removing redundant data from the dataset presented in Table 2. Hypothetically, as the faults are imitated in axis 4 of the robot, there is a risk that the current signals of the other axis motors may be affected, which would entail simultaneous MCSA concentrating on the data of all axes at the same time to identify the faults. However, this was not the case with the Hyundai Robot, as it was found that the fault in one axis would not affect the operation or signal pattern of the other axis motors. This gives the benefit of relying on the data on the only axis where the fault lies. Table 3 illustrates the size of the data after pre-processing and signal analysis. This dataset consists of a scalogram representation, rather than a signal cycle. Each signal cycle is translated to a scalogram representation, and the number of samples is reduced from 30/27/24 to 20/17/14 for the Normal/Faulty/Faulty Age class, respectively, due to the DQ0 Transformation.

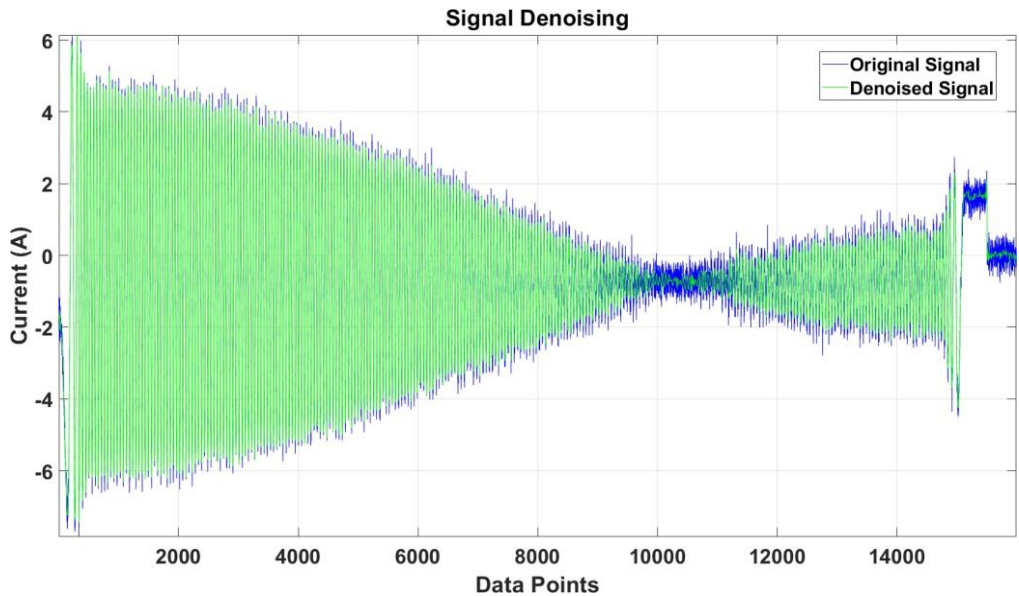

Figure 11: An example of the de-noised signal cycle after DWT implementation.

Figure 13 shows the detailed scalogram obtained after the CWT implementation. The exemplary scalogram is for the speed of 10%. The x-axis shows the time in milliseconds and the y-axis shows the scales chosen based on the frequency spectrum of the input signal. The colour bar on the right-hand side shows the magnitude of a specific scale at a specific instance. Note that the x- and y-axis values for each speed case were kept similar for better comparison purposes. Figure 14 shows the scalogram images for each speed domain for a single cycle. There is a clear difference in the scalogram images: as the speed of the rotation increases, the frequency also increases with different time scales. Figure 15 shows the comparison among three classes in an RGB and Greyscale format at a speed of 10%.

### 2.2.2. *Domain Knowledge Transfer*

The original ISMD dataset, which had several speed domains with scarce and imbalanced characteristics, could not be considered as a dataset to perform well for the classification of faults; and if the classification was carried out at a fair stage, the system would still be limited to an application-specific environment, with little interpretability capabilities. Therefore, it is necessary to either collect further data with thousands of samples, which are, in most cases, very complex for industrial robots and systems, or to adapt to a method in which data features with minimal samples across several domains can be translated, and based on those translated features, create a new dataset of translated knowledge, which can be further used to establish DL-based classification models. The latter was not feasible until recently when these studies [38, 39] suggested methods for image-to-image translation, where a GAN with an input source and the reference image is used to learn the features of the source and reference, generating a new image with features from both the source and the reference. GAN has shown positive results in the field of computer vision, being particularly efficient in areas such as image synthesis [47–49], image translation [50–52], and image creation [53–56]. The previous GAN image-to-image translation framework was limited to a single domain, where a single domain could be used with a single network. To translate images through multiple domains, a new network was required, increasing the computational cost for the generation of systems with the multi-domain image-to-image translation.

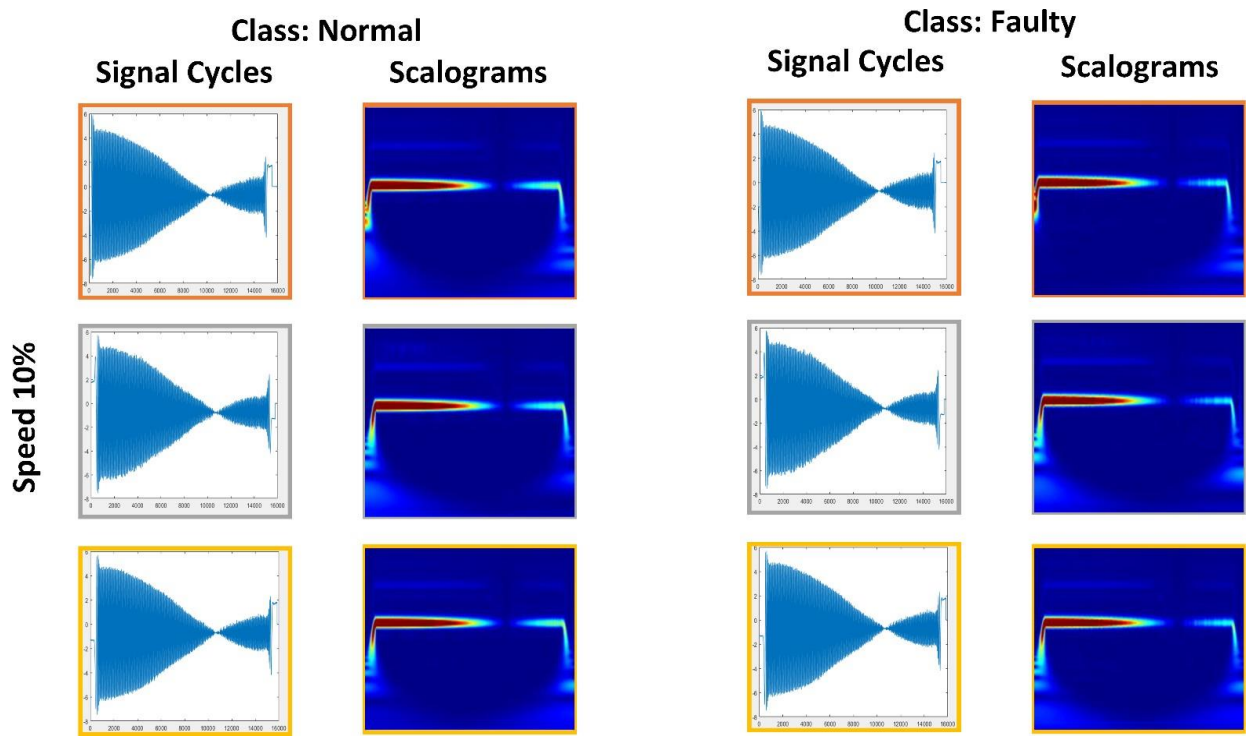

Figure 12: Scalogram images after CWT implementation for a single domain of 10 % rotation speed.

Table 3: Original ISMD data.

| Refined Original ISMD Dataset | Multi-Domain (Speed Profiles) | | | | | | | | | |
|---|---|---|---|---|---|---|---|---|---|---|
| | 10% | 20% | 30% | 40% | 50% | 60% | 70% | 80% | 90% | 100% |
| | Class; Normal \Faulty \ Faulty Age, Number of Samples (cycles) | | | | | | | | | |
| Total No. of Samples (Axis 4) | 20 \ 17 \ 14 | 20 \ 17 \ 14 | 20 \ 17 \ 14 | 20 \ 17 \ 14 | 20 \ 17 \ 14 | 20 \ 17 \ 14 | 20 \ 17 \ 14 | 20 \ 17 \ 14 | 20 \ 17 \ 14 | 20 \ 17 \ 14 |
| Dataset Size | 510 X 3 (No. of Samples X No. of Classes) | | | | | | | | | |

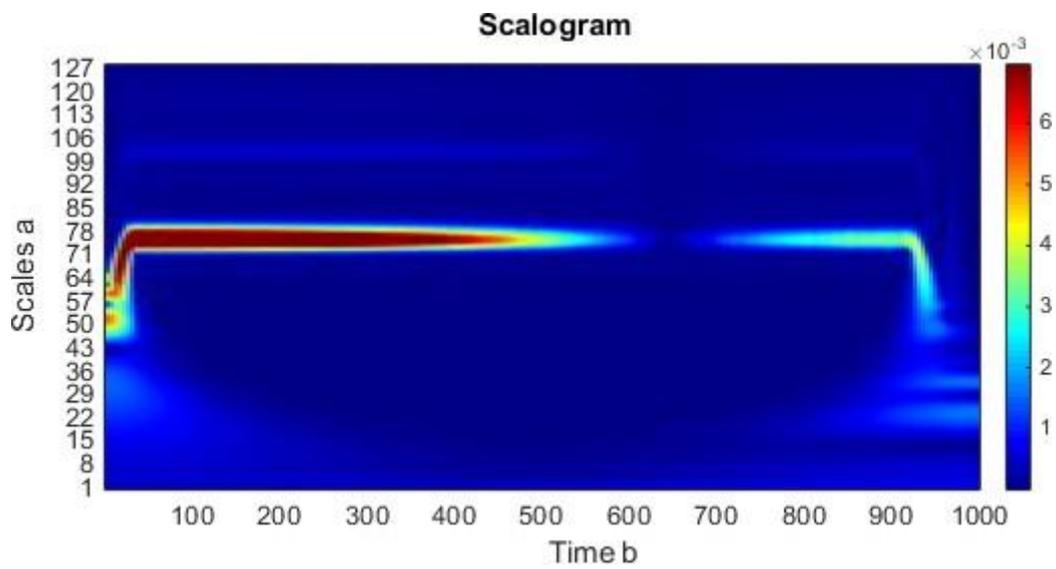

Figure 13: An example of a scalogram obtained after CWT implementation for the speed of 10%.

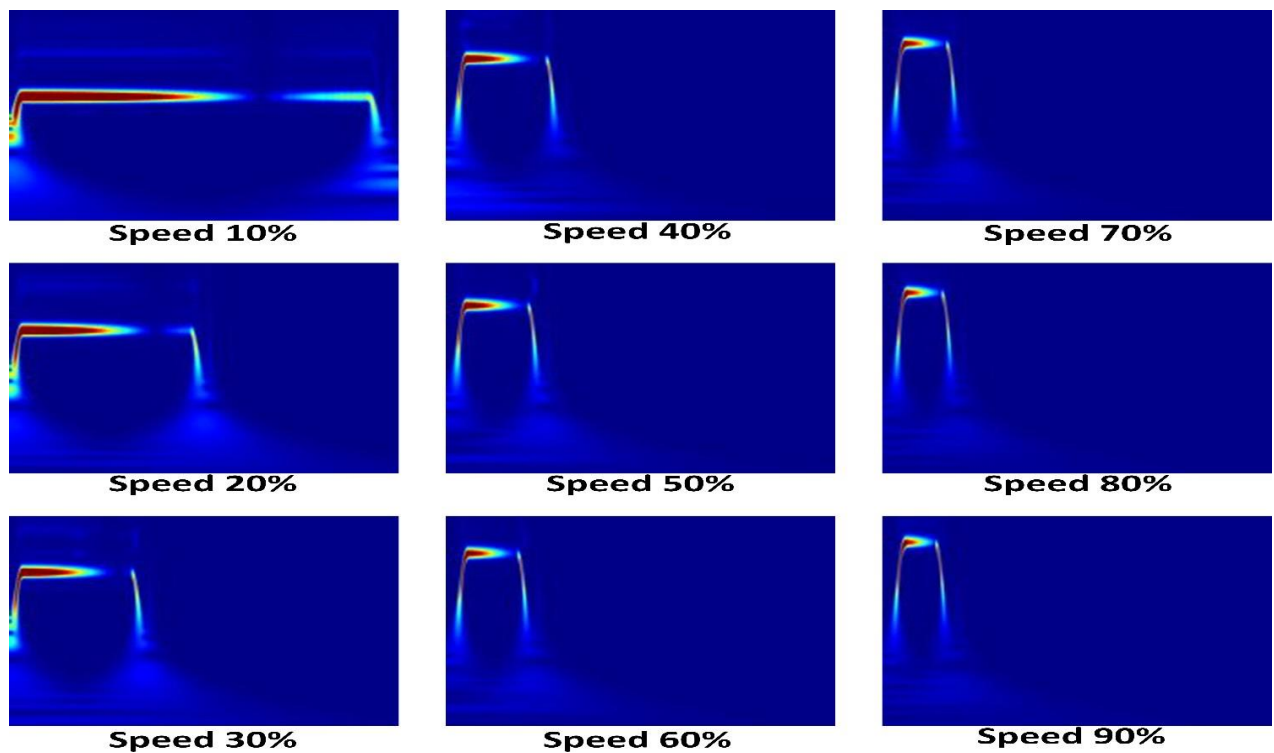

Figure 14: Scalogram images for multiple speed domains for a single cycle.

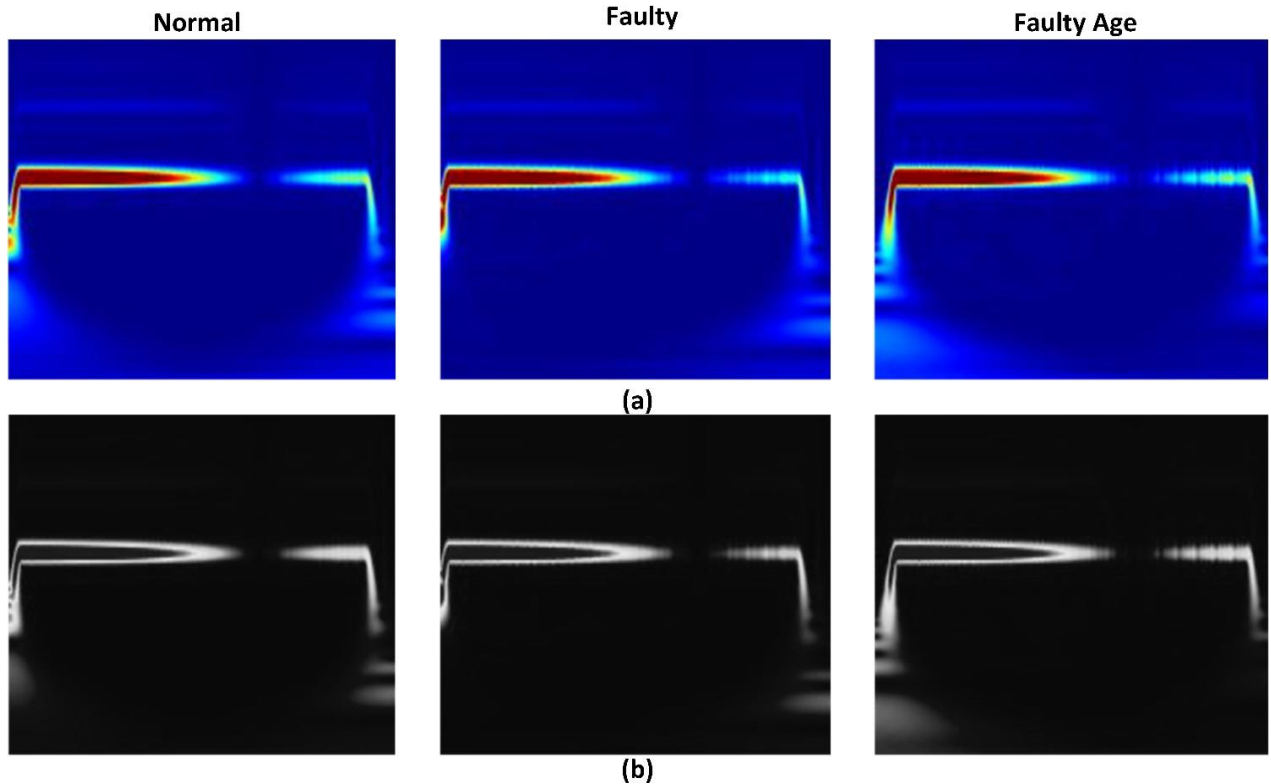

Figure 15: Comparison among different classes; (a) RGB images, and (b) Greyscale images.

Figure 16 demonstrates the difference between the previous cross-domain models and the StarGAN [38] used in this work, (a) addressing how twelve independent generator networks are expected to be trained to interpret images from only four various areas, where, as in (b), the model demands training data from several environments, and learns to map among all domains using only one generator. Figure 17 displays the StarGAN architecture. The first part of the network is a generator that takes an input image and converts it into an output image with a domain-specific style code. The second part is a mapping network that converts hidden code to style codes for various domains, which have been randomly chosen during the training process. The third part is a style encoder that derives an image style code that enables the generator to execute a reference-guided image fusion. The last part, like other GANs, is the discriminator that differentiates between original and fake images from a variety of domains.

### 2.2.3. Data Split and Classification

Using the domain knowledge transfer with StarGAN mentioned in the previous section, a dataset comprising of scalogram images from each class with translated images across each speed domain was created. Initially, three datasets for Normal, Faulty, and Faulty Age classes were generated using three StarGAN's trained specifically to tackle a single class with multi-domain data. The generated dataset with the original ISMD dataset was used

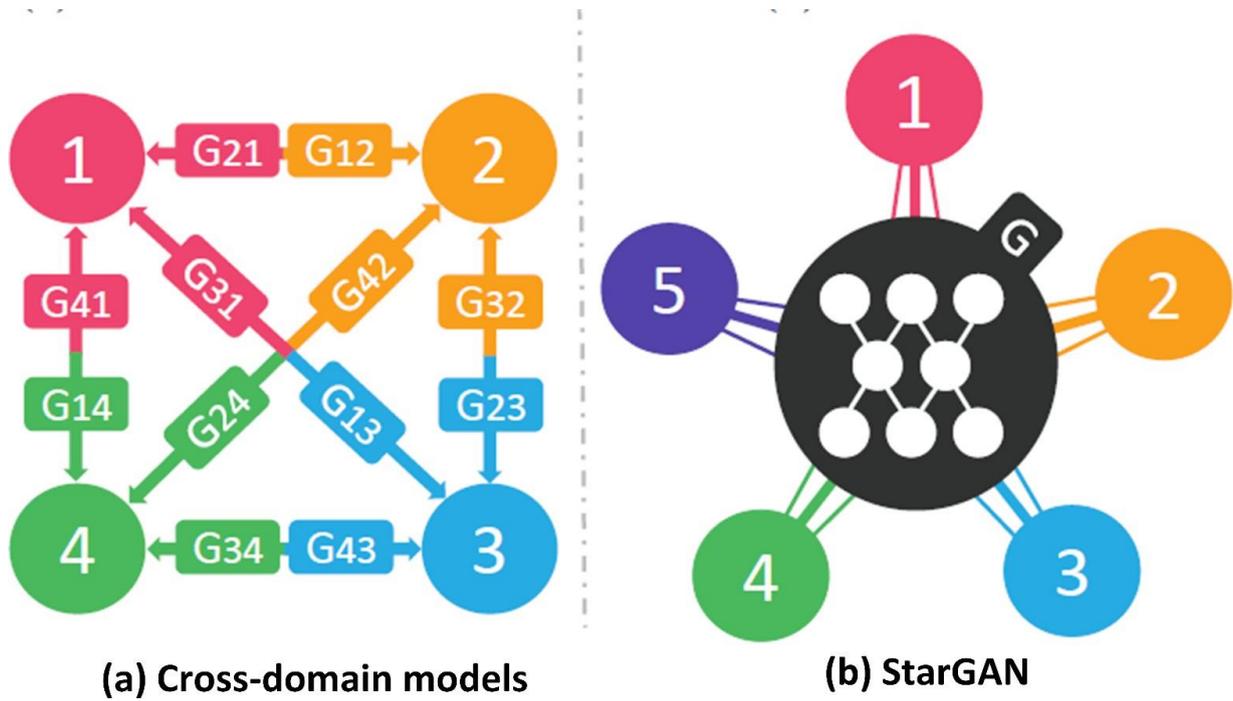

**(a) Cross-domain models**

**(b) StarGAN**

Figure 16: Comparison between (a) previous cross-domain models, and (b) StarGAN [39].

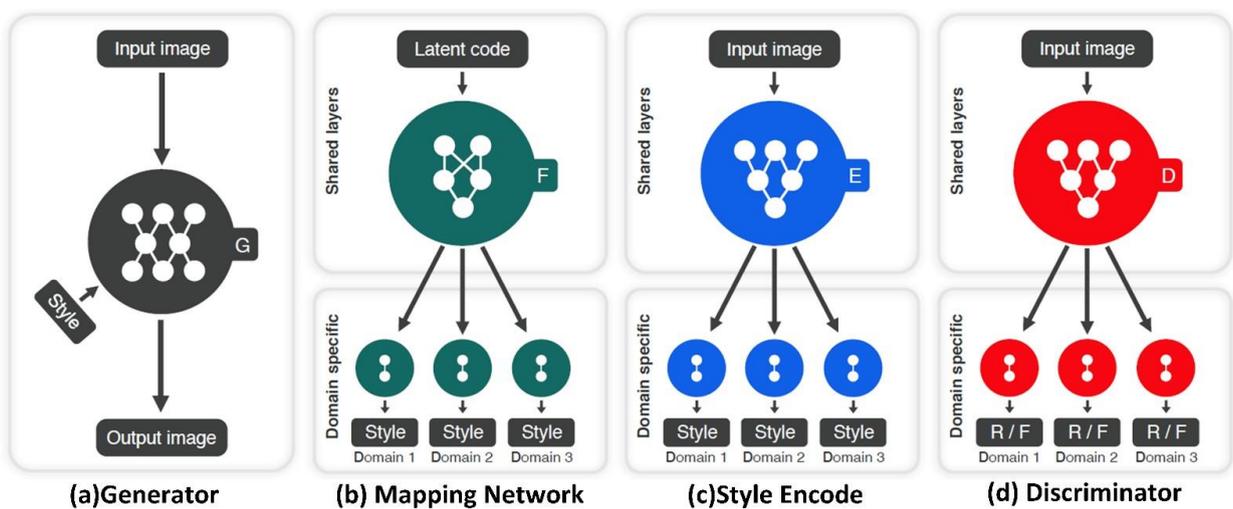

**(a)Generator**

**(b) Mapping Network**

**(c)Style Encode**

**(d) Discriminator**

Figure 17: Architecture of the StarGAN: (a) Generator, (b) Mapping network, (c) Style encoder, and (d) Discriminator [38].

as a final dataset where it was split into training and validation datasets. Figure 18 shows the details of the data generation and domain knowledge transfer for a single domain of 10% speed. Note that the digits on blocks show the total number of images present at the specific stage. The size of the generated images is 224 x 224 pixels. The same operation was carried out for each domain, and the final dataset was obtained. The total number of images for domain knowledge transfer among two domains can be calculated using Eq. (3):

$$N_{ab} = (n_a \times n_b) \times (n_d - N_d) \tag{3}$$

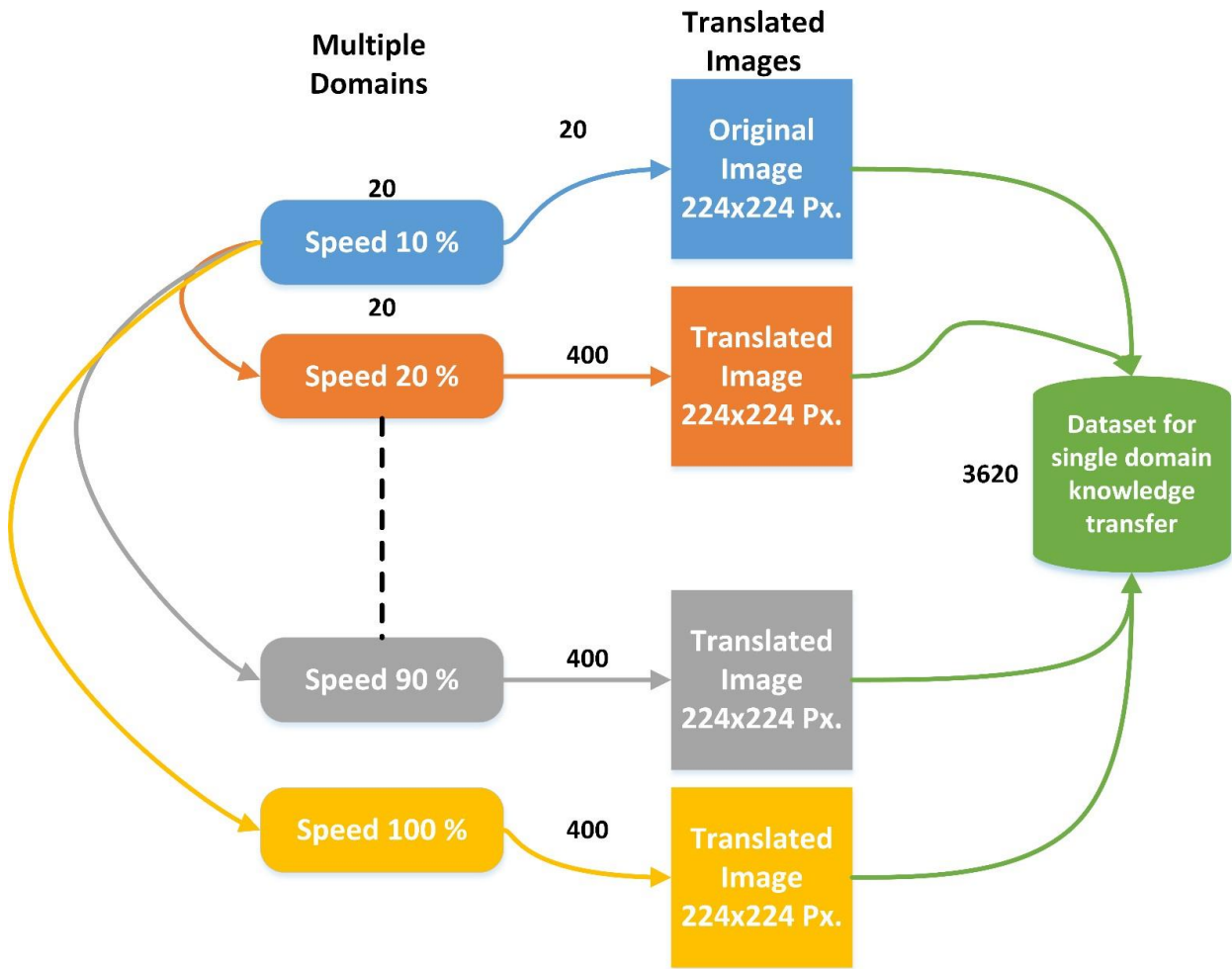

Figure 18: Detail of domain knowledge transfer and data generation for a single domain of 10 % speed.

Where $N_{ab}$ is the total number of images after knowledge transfer between domains $a$ and $b$, $n_a$ is the number of images in the domain $a$, $n_b$ is the number of images in domain $b$, $n_d$ is the total number of domains, and $N_d$ is the domain number = 1,2,3,. . . ,10 for speed (10,20,30,. . . ,100) %. Table 4 shows the possible combinations of domain knowledge transfer between different domains with the total number of the generated images (calculated using Eq. (3)), and the overall final dataset size. Note that domain 10 with 100 % speed is not present in the following table, since it has already been compensated in the data generation for the other domains.

The final dataset mentioned in Table 4 was used for the classification of the faults using transfer learning on some of the prominent benchmark CNN models. In this work, fine-tuning, a transfer learning principle that substitutes the pre-trained output layer with a layer containing the number of final dataset classes was used. The last three layers were substituted with a fully connected layer, a softmax layer, and a classification layer. The primary objective of practising pre-trained CNN models is to achieve quick and accurate training for CNN's manipulation of random initialization of weights and to accomplish a low training error. The effectiveness of the architectures of GoogLeNet [57], SqueezeNet [58], AlexNet [59], VGG16 [60], Inception v3 [61], and ResNet50 [62] has been analyzed for this specific fault detection and diagnosis system. Table 5 presents the characteristics of these CNN architectures.

## 3. Results and Discussion

### 3.1. Domain Knowledge Transfer for ISMD Data/Data Generation

Table 6 shows the specifications of the PC used in this work. For domain knowledge transfer, StarGAN was trained for 100,000 iterations. The recorded training time was approximately 5 hours for a single class and 15 hours for three classes. The number of iterations was carefully selected after observing the loss for the generator and the descriptor module of the StarGAN. Figure 19 shows some results of the latent images generated during the training process at the iteration of 10,000, 20,000, 30,000, and 40,000. Observably, at different iterations, the network learns more and more features, generating images with a similar pattern as the input. For precise analysis, results are presented with a focus on only one class: Normal. The results for the other classes were in a similar pattern, but with different features and information. In Fig. 19, (a) is for the speed domain of 30%, (b) 50%, (c) 80%, and (d) 100%.

Figure 20 shows some of the reference images generated during the training process of the network for different domains at the different number of iterations for the Normal class. The red rectangular box in Fig. 20 (a) represents the scalogram image for domain 1 that is the speed of 40 %, while the orange one represents domain 2 at 20 % speed, whereas the yellow one is the generated output image after domain knowledge transfer at the training iteration of 10,000. Figure 20 (b) shows the domain knowledge transfer between (10 and 30) % speed at the training iterations of 20,000, (c) (100 and 60) % speed at the training iterations of 30,000, and (d) (80 and 100) % speed at the training iterations of 40,000.

Figure 21 shows the results of the generated images after training the network. The reference images are the images provided to the network as a reference, with source images acting as the input at a specific time. The network takes the source image, compares it with the reference image, and based on the learned features from both sources and references, generates the output image. For multiple speed domains, 10 to 100 %, a relationship is established, and several images are generated based on the combinations previously presented in Table 4. Note that there are combinations in Fig. 21, such as (10,10), (20,20), (30,30), (40,40), (50,50), (60,60), (70,70), (80,80), (90,90), and (100,100), which act as Don't Care conditions (similar as in Table 4). These combinations are segregated from the final dataset since the goal is to transfer knowledge across multiple domains with the infusion of several frequency- and time-domain features at different levels of speed, rather than the same set of speed domains. This also reduces the size and redundancy of the final dataset. As previously mentioned, similar images as in Fig. 21 are generated for the other two classes: Faulty and Faulty Age, to create a complete dataset. The implementation of the StarGAN is available as an opensource for both the PyTorch and Tensorflow libraries at the GitHub repository [63]. In this particular work, the TensorFlow implementation was used.

Table 4: The final dataset with possible combinations of domain knowledge

| Number of combi-nition/opera-tions | Domain 1 (10% Speed) | Domain 2 (20% Speed) | Domain 3 (13% Speed) | Domain 4 (40% Speed) | Domain 5 (50% Speed) | Domain 6 (60% Speed) | Domain 7 (70% Speed) | Domain 8 (80% Speed) | Domain 9 (90% Speed) |
|---|---|---|---|---|---|---|---|---|---|
| 1 | (10,20) | x | x | x | x | x | x | x | x |
| 2 | (10,30) | (20,30) | x | x | x | x | x | x | x |
| 3 | (10,40) | (20,40) | (30,40) | x | x | x | x | x | x |
| 4 | (10,50) | (20,50) | (30,50) | (40,50) | x | x | x | x | x |
| 5 | (10,60) | (20,60) | (30,60) | (40,60) | (50,60) | x | x | x | x |
| 6 | (10,70) | (20,70) | (30,70) | (40,70) | (50,70) | (60,70) | x | x | x |
| 7 | (10,80) | (20,80) | (30,80) | (40,80) | (50,80) | (60,80) | (70,80) | x | x |
| 8 | (10,90) | (20,90) | (30,90) | (40,90) | (50,90) | (60,90) | (70,90) | (80,90) | x |
| 9 | (10,100) | (20,100) | (30,100) | (40,100) | (50,100) | (60,100) | (70,100) | (80,100) | (90,100) |
| Generated Im-ages/class | 3620 | 3220 | 2820 | 2420 | 2020 | 1620 | 1220 | 820 | 420 |
| Final Datset Size | 40847 x 3 (No. of Samples X No. of Classes) | | | | | | | | |

Table 5: Characteristics of the benchmark CNN models.

| Network | Depth / No. of Layers | Input Image Size |
|---------|----------------------|------------------|
| GoogLenet | 22 | 224 x 224 |
| SqueezeNet | 18 | 227 x 227 |
| AlexNet | 8 | 227 x 227 |
| VGG16 | 16 | 224 x 224 |
| Inceptionv3 | 48 | 299 x 299 |
| ResNet50 | 50 | 224 x 224 |

Table 6: PC specifications.

| Component | Detail |
|-----------|--------|
| CPU | Intel® Core™ i7-8700k CPU Eight-core @ 3.0 GHz |
| Memory | 32 GB |
| GPU | NVIDIA GeForce RTX 2080 Ti |
| Operating System | Linux |

The dataset obtained from the StarGAN with the original dataset recorded in real-time was used by keeping the resource parameters intact. Each class of the generated dataset with the corresponding class of the recorded dataset was used for training and validation purposes. The size of the final dataset (total dataset) was 40,847 ✗ 3 (Number of samples/images ✗ Number of classes).

## 3.2. Faults Classification

Performing the classification tasks of the faults and validating the efficacy of the proposed data generation methodology, transfer learning on some of the prominent benchmarks of CNN was adopted. Six CNN models where each having a different number of layers and input image size, each developed with a different analogy to carry out the image classification tasks were trained. For comparison purposes, each of these models with the same number of epochs with similar batch sizes and hyper-parameters was trained. Table 7 presents the values of hyper-parameters used during the training of the CNNs. The training was performed on the PC with the specifications mentioned in Table 6. The final generated dataset was used as a training dataset. The validation dataset contained the data from the original ISMD dataset, whereas the training dataset was composed of the data generated using StarGAN. This was done to verify the interpolation capacity of the trained CNN models and to avoid overfitting. The validation is performed on unseen, un-generated (data not generated by StarGAN) data. The training dataset was further divided into the training, testing, and validation dataset with the ratio of 70%:30%. The split strategy was chosen carefully to the said ratio to counter the computationally intensive task of classification.

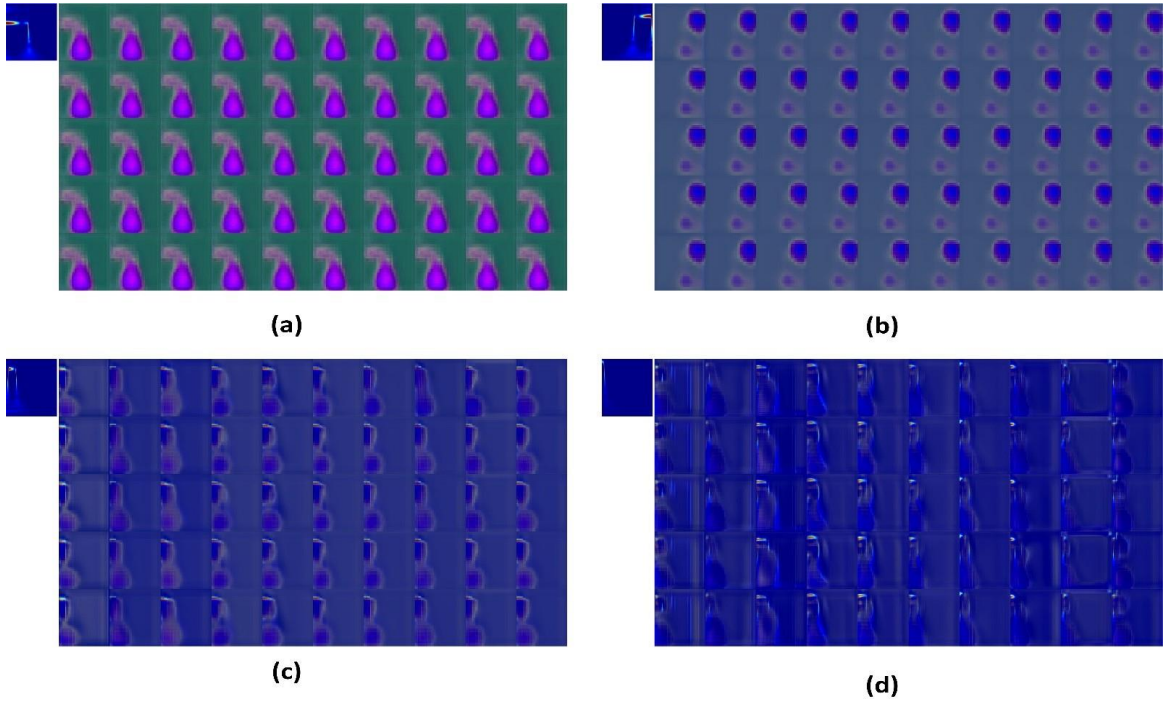

Figure 19: The latent images generated during the training process for class Normal:(a) at 10,000 iterations for 30 % speed, (b) 20,000 iterations 50 % speed, (c) 30,000 iterations for 80 % speed, and (d) 40,000 iterations for 100 % speed.

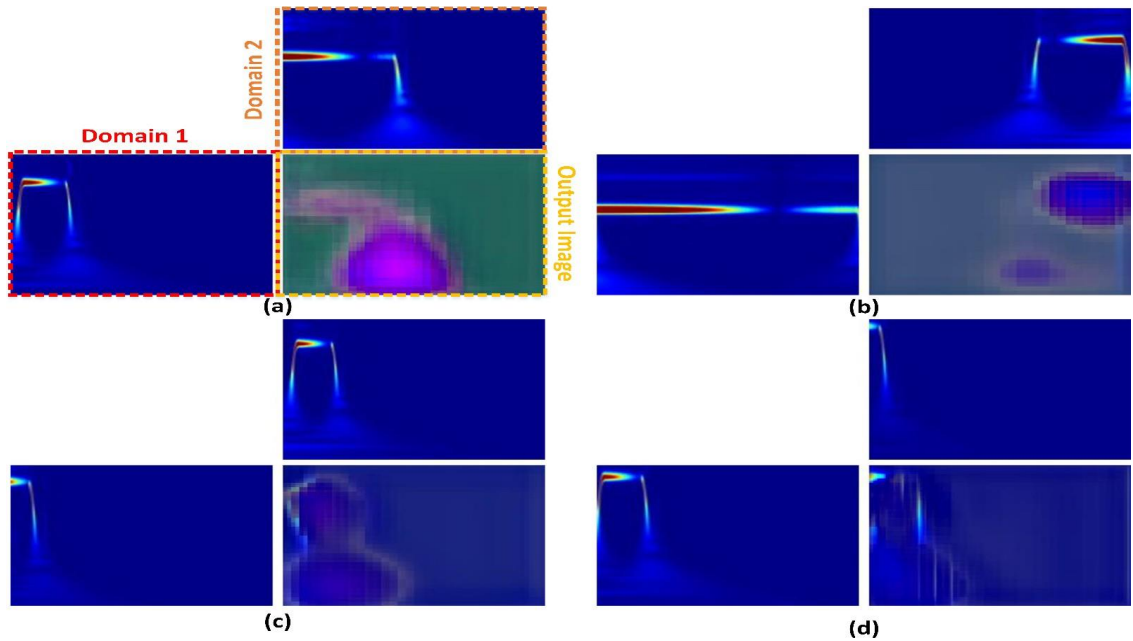

Figure 20: The reference images generated during the training process with domain knowledge transfer, (a) speed (40 and 20) % at 10,000 iterations, (b) (10 and 30) % at 20,000 iterations, (c) (100 and 60) % at 30,000 iterations, and (d) (80 and 100) % at 40,000 iterations.

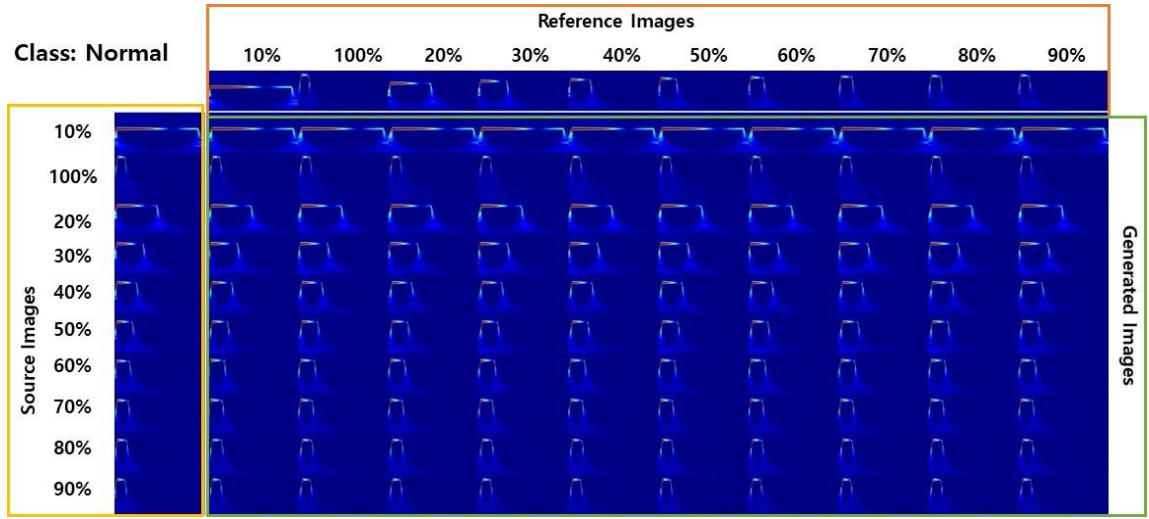

Figure 21: The generated images for the class Normal with domain knowledge transfer.

Table 7: Hyper-parameters used for training.

| Hyper-Parameters | Value |
| --- | --- |
| Batch Size | 32 |
| Epochs | 30 |
| Learning Rate | 0.0001 |
| L2 Regularization | 0.00001 |
| Optimization Algorithm | Stochastic Gradient Descent |

The performance of the six CNN models (GoogLeNet, SqueezeNet, AlexNet, VGG16, Inception v3, and ResNet50) was evaluated using the following performance metrics. Some of these metrics are widely used to test the performance of a trained ML/DL model. Among these metrics, Accuracy, Sensitivity, Specificity, Precision, and F-score were used. Accuracy is the most important metric, and it tells the number of samples that are correctly classified out of all the samples. Generally, it is expressed as the ratio of True Positives (TPs) and True Negatives (TNs) divided by the sum of the total number of TPs, TNs, False Positives (FPs), and False Negatives (FNs). A TP or TN is a data sample that the algorithm accurately classifies as true or false. On the other hand, an FP or FN is a data sample that the algorithm falsely classifies. Equation (4) represents this metric:

$$Accuracy = \frac{TP + TN}{TP + TN + FP + FN} \tag{4}$$

The sensitivity metric is also known as the recall. It is defined as the number of accurately classified positive samples, which implies that how many samples of the positive classes are identified correctly. It is given in the Eq. (5).

$$Sensitivity \ / \ Recall = \frac{TP}{TP + FN} \tag{5}$$

The specificity refers to the conditional possibility of the TN in the given class, this implies that the prediction of the possibility of a negative label becoming true. It can be represented as in Eq. (6).

$$Specificity = \frac{TN}{TN + FP} \tag{6}$$

Precision, calculated as the percentage of TP, divided by the number of TP plus the number of FP, is given by Eq. (7). This metric is about consistency, i.e., it measures the prediction performance of the algorithm. It tells us that how precise a model is out of what is expected to be positive, and how many of those are truly positive.

$$Precision = \frac{TP}{TP + FP} \tag{7}$$

Finally, the F-score, defined as the relative average of precision and recall, as shown in Eq. (8). It relies on positive class evaluation. The high value of this parameter suggests that the model performs the best in the positive class. Table 8 presents the results achieved using fine-tuning of the CNN's based on the performance metrics.

$$F - Score = 2 \times \left( \frac{Precision \ \times Recall}{Precision + Recall} \right) \tag{8}$$

The achieved results are significantly promising for all six CNN models, starting with GoogLeNet, which achieved an accuracy of 99.7% with high predictability for the positive and negative classes with a sensitivity of 99.6% and specificity of 99.7%. It was undoubtedly the best model to classify the faults. Following it was SqueezeNet with an accuracy, sensitivity, and specificity of 99.6%, 99.4%, and 99.6%, respectively. SqueezeNet performed considerably well for the overall classification. The third best was VGG16, which achieved the accuracy, sensitivity, and specificity of 99.3%, 99.0%, and 99.4%, respectively, followed by AlexNet, which showed less accurate results, compared to the top three models. The maximum accuracy was recorded as 98.0% with sensitivity and specificity of 97.2% and 98.4%, respectively. Inceptionv3 and ResNet50 performed considerably less effective than the other models but were still better at solving the fault classification problem.

Table 8: Results of the evaluation metrics and CNN performance.

| CNN model | Metrics | | | | |
| | Accuracy | Sensitivity | Specificity | Precision | F-score |
|---|---|---|---|---|---|
| **GoogLeNet** | 99.71325864 | 99.61988796 | 99.75994398 | 99.61988796 | 99.619888 |
| **SqueezeNet** | 99.61988796 | 99.47983193 | 99.68991597 | 99.48506224 | 99.4798152 |
| **VGG16** | 99.3331 | 99.0997 | 99.4499 | 99.102 | 99.1 |
| **AlexNet** | 98.01269841 | 97.21904762 | 98.40952381 | 97.37777778 | 97.220979 |
| **Inceptionv3** | 97.9994 | 97.2992 | 98.3496 | 97.366 | 97.298 |
| **ResNet50** | 95.7055 | 94.2583 | 96.4921 | 94.7586 | 94.2398 |

The architectures of these CNN models are different from one another. GoogLeNet is composed of 22 layers, where there are multiple convolutional layers, followed by the inception module, and more convolutional layers. The greater numbers of layers help the network extract useful and distinguishable prominent features from a given training dataset. The model is heavy compared to the others but performs well in learning features with multiple filters. On the other hand, SqueezeNet is composed of 18 layers, designed with a different analogy from GoogLeNet, and has relatively fewer parameters. VGG 16 and AlexNet are composed of 16 and 8 layers, respectively, where convolutional layers are stacked one after another. While they were proven to be promising models for the problem of image classification, in this study, they were inefficient for the problem of fault classification. While the number of layers can be an important parameter to consider when designing the architecture of a neural network, the relationship among the different layers of the network can also play a significant role in improving the learning capacity of a network. Inceptionv3 and ResNet50 are examples of such networks. Even though both these networks have a high number of layers compared to GoogLeNet, they nevertheless do not perform well for the fault classification problem. Inceptionv3 is composed of 48 layers, while ResNet50 is composed of 50 layers. The poor performance of these architectures has to do with the way the layers are related to one another in the network. In Inceptionv3, multiple inception modules are used that are comprised of multiple convolutional layers; whereas, in ResNet50, the skip connection concept of residual networks is used, where the output from a layer can be directly fed as an input to the other layer at a specific level of the architecture while skipping the layers between them. The achieved results in this work also prove that such a network fails to show more prominent results, due to the design of the architecture, and specific implementation methodologies.

Figure 22 shows the confusion matrix for the top two models. The rows correspond to the expected output class, while the columns correspond to the real target class. The diagonal cells correspond to the groups that are precisely classified. Off-diagonal cells correspond to groups that are falsely classified. Each cell displays the number of samples and their percentile. The far-right column represents the percentiles of all the samples expected for each class, correctly and incorrectly classified. These metrics are also known as the precision and the false discovery rate. In both cases of GoogLeNet and SqueezeNet, more confusion is found between the Normal and Faulty classes. Figure 15, which shows the scalogram images of each class, also confirms this. The difference between the Normal and Faulty Age class is more visible than the Faulty class, making it easier for the CNN model to classify with high accuracy. Almost 100% accuracy between the Normal/Faulty Age and Faulty/Faulty Age class is achieved.

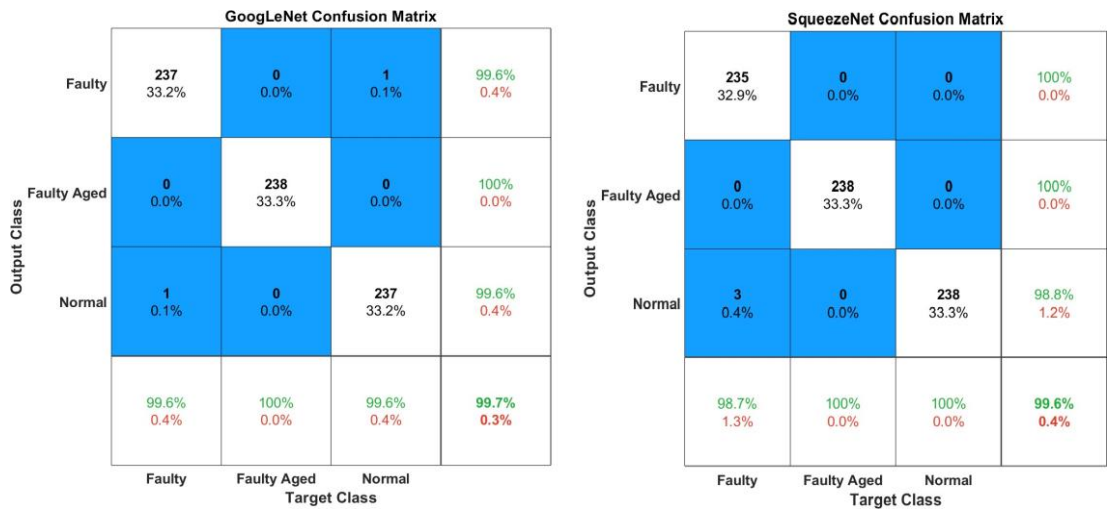

Figure 22: Confusion matrix for GoogLeNet and SqueezeNet.

## 4. Conclusion

This study provides a holistic approach to the detection and diagnosis of faults related to the mechanical component of an industrial robot in ISMD data settings for the component level PHM. The obtained results were impressive for six CNN benchmark models: GoogLeNet, SqueezeNet, VGG16, AlexNet, Inceptionv3, and ResNet50, with accuracies of 99.7%, 99.6%, 99.3%, 98.0%, 97.9%, and 95.7%, respectively. The results achieved demonstrate that the suggested methodology can overcome the key problem of ISMD data. The benefit will be the ability to design potential methods for detecting and diagnosing faults beforehand, with less knowledge about the type of fault. Furthermore, interpolation between various types of faults and different robots will be feasible, due to the transition of domain knowledge. By merely collecting or providing a limited amount of data, it could be used to construct DL models that are highly efficient, and that can be applied in real-time with a realistic framework. For future work, there is a need to concentrate on the implementation of further faults related to electrical and mechanical components in various types of robots to examine and build a system that can recognize not just faults but also the robots where the fault has occurred.

**Funding:** This research was financially supported by the Ministry of Trade, Industry, and Energy (MOTIE) and the Korea Institute for Advancement of Technology (KIAT) through the International Cooperative R&D program (Project No. P059500003)

**Acknowledgments:** The author is grateful for the support provided by the team of SMD Lab, Dongguk University, South Korea including Professor Heung-soo Kim, Hyewon Lee, and Izaz Rauf in helping with communication with industry partner and providing with the resources and funding for the project.

**Conflicts of Interest:** The author declares no conflict of interest.